\documentclass[10pt,twocolumn]{article}          
\usepackage{latex8}                              
\usepackage{epsfig}                               

\begin{document}


\title{Parametric Estimation of the Ultimate Size of
Hypercomputers}
\author{Dmitry Zinoviev\\
Mathematics and Computer Science Department\\ 
Suffolk University, Boston, 02114 USA\\
dmitry@mcssuffolk.org\\\ }
\date{}
\maketitle
\thispagestyle{empty}

\begin{abstract}
The performance of the emerging peta\-flops-scale supercomputers of the
nearest future (hypercomputers) will be governed not only by the clock
frequency of the processing nodes or by the width of the system bus,
but also by such factors as the overall power consumption and the
geometric size. In this paper, we study the influence of such
parameters on one of the most important characteristics of a general
purpose computer --- on the degree of multithreading that must be
present in an application to make the use of the hypercomputer
justifiable. Our major finding is that for the class of applications
with purely random memory access patterns ``super-fast computing''
and ``high-performance computing'' are essentially synonyms for
``massively-parallel computing''.
\end{abstract}

\Section{Introduction}

Super-fast computers processing data at a sustained rate on the order
of 10$^{15}$ integer or floating-point operations per second (1
petaops, or 1 petaflops), also known as
hypercomputers~\cite{sterling99}, will be emerging within the next
decade as ultimate tools for solving very large-scale problems of
computational fluid dynamics, weather forecasting, nuclear stockpile
stewardship, cryptanalysis, real-time image processing and rendering,
and the like~\cite{gao96}.

Common sense supported by the results of preliminary case
studies~\cite{wittie98} suggests that the hypercomputers will
materialize as hardware installations of substantial size and power
consumption. The average geometric diameter of the installation,
combined with the ultra-high clock frequency, will be eventually
translated into a memory access latency of several hundreds and
thousands processor cycles, --- a situation unthinkable in the domain
of personal computers but quite common on the Internet. To achieve and
sustain the required performance, the hypercomputer must be originally
designed as a highly multithreaded machine~\cite{tera97,gao97}.
Preemptive multithreading helps to hide the memory access latency.
However, it implies high parallelism, which inevitably limits the
usability of a hypercomputer to a narrow domain of intrinsically
parallel applications.  Careful consideration of physical factors can
help to anticipate the potential problems that may render the design
of a hypercomputer doomed to failure.

In this paper, we will obtain a rough parametric estimation of the
performance of hypercomputers based on their fundamental physical and
geometric properties, such as power consumption and wire size.

\Section{\label{sec:model}Model}

For the purpose of this study, the following simplified model of a
hypercomputer has been used.  We assume that the hypercomputer
consists of $Q$ nodes, each node being either a processing element
(PE), or a memory bank. The nodes are connected using a multistage
internal network. The diameter of the network $D$ is on the order of
$\log_2Q$ (this is true for delta networks and approximately true for
other high-performance networks). For the ease of application development,
all processing elements have uniform access to the globally shared
memory. A typical application using the hypercomputer generates purely
random memory traffic at a rate of 1.32 (``load'') requests and
0.78 replies (``store'') per clock cycle~\cite{patterson96}, or
approximately 1 outbound message per cycle per node. All instructions
are presumed to be fetched from local instruction caches and do not
contribute to the total traffic.  Data caches are not considered,
taking into account the random pattern of the memory usage. Finally,
we assume that the processor word width is $W$ bits, the processor
clock frequency is $f_0$, and that each PE completes one instruction
per clock cycle.

To achieve its ultimate performance, the system must be well balanced
in a sense that the round-trip memory access latency, measured in
PE clock cycles, should be approximately equal to the degree of
multithreading. In this case, a thread blocked at a memory request
will be scheduled for execution by the hardware exactly when the results
arrive to the local registers. Smaller degree of multithreading will
reduce the performance of the hypercomputer, while higher degree will
require extra hardware for thread contexts, most of
which will never be used.

\begin{figure}[!bt]\centering
\epsfig{file=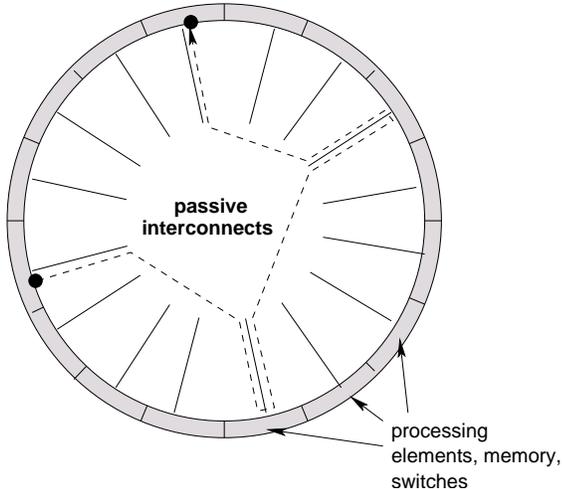,width=0.9\columnwidth}
\caption{\label{pict:schema}Arrangement of the components of a
hypercomputer. A sample message path is shown with a dashed line.}
\end{figure}

As a first step toward the refining of the proposed model, we observe
that the design of a petaflops-scale hypercomputer implies
three-dimensional integration. Indeed, it has been
shown~\cite{dorojevets98c} that the footprint of a hypercomputer
flattened into the two-dimensional space would be as large as a soccer
field (namely, $\sim1,000\,m^2$). The actual arrangement of the
components (PEs, memory banks, and internal network nodes) is not
essential for the study.  We will focus on a rather unrealistic, but
easy to model, spherical configuration, with all active components
evenly placed on
the surface of a sphere of diameter $L$, and all passive components
(wires) hidden under the surface, as shown in
Figure~\ref{pict:schema}. (A similar --- but technically more
sound --- cylindrical arrangement has been proposed
in~\cite{abelson98} and \cite{wittie98a}.) Such configuration permits
relatively
easy access to the active components in case they need maintenance or
replacement.

\Section{Power Consumption}

Electrical power is consumed by the hypercomputer statically and
dynamically. The static term is attributed to the leakage current
(which can be ignored, at least in theory) and to the power
dissipation in passive interconnecting wires. The dynamic term depends
on the performance $\Theta$ of the hypercomputer.

Let us begin with the evaluation of the total number of wires required to
interconnect the processing elements. The signal transfer rate on a
wire (limited by the wire bandwidth $B_w$) may be substantially slower
than the PE clock rate $f_0$. Respectively, the amount of passive
wires in the network must be proportionally larger to match the total
bandwidth $B$ of requests generated by the PEs, and the available
bandwidth of the network. Each stage of a multistage network
contributes proportionally to the total number of wires, too. Finally,
we must add extra wires to compensate for the network saturation,
which typically takes place at $\alpha\sim60\,\%$ load:

\begin{equation}
N=\frac{B}{B_w}\frac{D}{\alpha} =
\frac{\left(1.1f_0WQ\right)D}{B_w\alpha} \sim
\frac{f_0WQD}{B_w\alpha}. \label{eq:N}
\end{equation}

Second, we must establish a relationship between the performance of
the hypercomputer and its configuration and clock frequency. The
aggregate peak performance of the hypercomputer, measured in floating
point operations per second, can be roughly estimated as
\begin{equation}
\Theta=Qf_0\left(W/W_0\right),\label{eq:X}
\end{equation}
where $W_0$ is the number of bits per word in a ``standard''
processing element. The ratio in the parentheses takes into account
the fact that, for instance, a 128-bit PE is twice as powerful as its
64-bit counterpart running at the same clock rate.

The size of the hypercomputer installation will be defined by the
amount of power $p_v$ that can be possibly removed from a unit
volume by means of either forced-air or water cooling. The maximum
power that can be removed in the former case is
$p_s=5\cdot10^5\,W/m^2$~\cite{itrs99}. Water cooling can remove more
power, but requires more sophisticated and bulky plumbing. At the
moment, we do not know what will be the ultimate vertical chip pitch
$h$ for the 3-dimensional integration. The pitch of $h=5\,mm$ sounds
like a sane approximation, with a proper allowance for the packaging
and cooling infrastructure. Under this assumption, the maximum power
that can be removed from a unit volume is $p_v=p_s/h=10^8\,W/m^3$.

\SubSection{\label{ssec:testvehicle}``Test Vehicle'' Hypercomputer}

To verify our theoretical reasonings, we will consider a hypothetical
hypercomputer of year 2007. This ``Test Vehicle'' hypercomputer (TVHC)
will
be driven by $Q=50,000$ super-fast 128-bit Intel chips
($f_0=20\,GHz$~\cite{paulson01}). The nodes will be connected using a
banyan network ($D=\log_2Q\approx16$) implemented as a collection of
insulated thin pure copper wires (bandwidth per wire
$B_w\approx3.6\,Gbps$~\cite{itrs99}; resistivity $\rho=17.5\cdot
10^{-9}\,\Omega\cdot m$; wire electrical cross-section
$\sigma_w=2.5\cdot10^{-8}\,m^2$). One can verify using Eq.~\ref{eq:X}
that the peak performance of this hypercomputer will be $10^{15}$
operations per second, or 1 petaops.

\SubSection{Static Power Dissipation}

Power dissipated statically by a passive resistive electrical system is
given by Ohm's law: $P_s=I^2R$, where $I$ is the signal current, and $R$
is the overall resistance of the system. We assume that
$I\approx\pm20\,mA$, although higher-current drivers may be needed to
sustain error-prone high bit rate transmission at meter-scale
distances.

The interconnection network can be ultimately considered as a
collection of $N$ individual wires of length $l_i$, with electrical cross-section
$\sigma_w$, made out of a good conductor with resistivity $\rho$. It
can be shown that the average distance between any two components on a
sphere $\bar L$ is $2L/\pi$. The wires are connected in
series, and the total resistance is:
$$
R=\sum_i^N R_i=\frac{\rho}{\sigma_w}\sum_i^N l_i=
\frac{\rho N\bar L_{st}}{\sigma_w}=
\frac{2\rho NL_{st}}{\pi\sigma_w}
$$
Finally,

\begin{equation}
P_s=\frac{2I^2L_{st}\rho N}{\pi\sigma_w}.\label{eq:static}
\end{equation}

The total heat generated by the static power dissipation $P_s$ must be
removed from the chip, according to the conditions stated above. This
is only possible, if the volume occupied by the wiring is large enough:
$P_s\le p_vV=\pi p_vL_{st}^3/6$. Substituting $P_s$ from Eq.~\ref{eq:static} and
$N$ from Eq.~\ref{eq:N} and Eq.~\ref{eq:X}, we get the final dependency of
$L_{st}$ on $\Theta$:

\begin{equation}
L_{st}\approx\sqrt{\Theta W_0\left(\frac{\rho I^2}{\sigma_w
p_v B_w}\frac{D}{\alpha}\right)}.\label{eq:Lstatic}
\end{equation}

The diameter of the ``static thermal core'' for the TVHC $L_{st}$ is $\approx0.008\,m$.

\SubSection{Dynamic Power Dissipation}

Dynamic power dissipation is due to the fact that each operation
executed by any PE requires certain energy (in our case,
$w\approx 10^{-10}\,J/op$~\cite{itrs99}). We have to consider heat
generated by both processing elements and memories (there are $2Q$ of
them), and switching elements (there are at least $QD/2$ of them,
assuming a delta-class interconnection network). We do not know the
exact relationship between the complexity of operations executed by
the switching engines and computational engines, and for the purpose
of this study we will assume that they are equivalent. Therefore, the
total dynamic power dissipation $P_d$ in the hypercomputer is equal to
$\Theta w\left(2+D/2\right)$. According to the model proposed in
Sec.~\ref{sec:model}, active processing and switching elements are spread on the
surface of the sphere enclosing the passive interconnection wires,
forming an ``active shell''. The surface of the sphere must be
spacious enough to enable adequate heat removal:
$\Theta w\left(2+D/2\right)\le\pi L_{dyn}^2w_0$. Obviously,

\begin{equation}
L_{dyn}=\sqrt{\frac{\Theta w}{\pi w_0}\left(2+\frac{D}{2}\right)}.\label{eq:Ldyn}
\end{equation}

For the TVHC, $L_{dyn}=0.8\,m$. This is certainly an optimistic
estimation, because a lot of power is required for various support
operations, such as PE ``housekeeping'' and memory refreshing.

\SubSection{Power Dissipation in Drivers}

Yet another source of dynamic power consumption is the set of drivers
responsible for the transmission of digital signals from one agent to
another along the interconnecting wires. Each driver constitutes a
current source injecting either $+I$ or $-I$ into the attached wire,
at voltage $V$. To reduce noise and decrease bit error rate, the drivers must be placed as close to the agents as
possible, and therefore are located on the same surface of the
``active core''. Altogether, $2N$ drivers are required, with
the total power dissipation of $P_{dr}=2NIU$. Again, the surface of the core
must be spacious enough:
$2NIU\le\pi L_{dr}^2p_s$. Naturally,
\begin{equation}
L_{dr}=\sqrt{\frac{2NIU}{\pi p_s}}=\sqrt{\Theta W_0}\sqrt{\frac{2IUD}{\pi
p_sB_w\alpha}}. \label{eq:Ldriver}
\end{equation}
Under the assumption of a really low-voltage driver ($V=1\,V$), the
diameter of the ``thermal core'' expands to $L_{dr}\approx4.9\,m$.

The diameters of all three thermal spheres considered so far ---
Eq.~\ref{eq:Lstatic}, Eq.~\ref{eq:Ldyn}, and Eq.~\ref{eq:Ldriver} --- scale as
$\sqrt{\Theta}$. This means, in particular, that the size of
the shell will be determined by static, dynamic, or driver-related
power dissipation, but not by all of them at a time. More
specifically,

\begin{equation}
L_{pow}=\max{\left(L_{st}, L_{dyn}, L_{dr}\right)}.\label{eq:Lpower}
\end{equation}

To summarize: the size of the ``minimal thermal core'' of the
hypercomputer suggested in Subsection~\ref{ssec:testvehicle} must conform to the
driver power dissipation requirements. The surface of the conforming
core will be large enough to accommodate the processing and switching elements, and
the volume of the core will be large enough to fit the interconnection
wires --- without introducing additional power constraints.

\Section{Wiring Constraints}

Alternatively, the ultimate size of a hypercomputer can be estimated
by considering how much space is required to contain the copper wires
constituting the interconnection network.

If a cross-section of a single interconnecting wire (including
appropriate insulation, cooling, mechanical support, etc.) is
$\sigma$, and there is the total of $N$ wires constituting the
interconnection network, then the total physical volume $V_1$ occupied
by the wiring is:
$$
V_1=\sigma N\bar L_g=2\sigma NL_g/\pi.
$$
On the other hand, this volume cannot exceed the volume of the core:
$$
V_2=\pi L_g^3/6.
$$
Therefore, the following simple equation holds: 
\begin{equation}
L_g=\sqrt{12\sigma N}/\pi\sim\sqrt{\sigma N}.\label{eq:L}
\end{equation}
For interchip connections implemented on a printed circuit board (PCB),
$\sigma$ may be chosen to be on the order of $10^{-7}\,m^2$ (wires are
placed at $\approx 0.3\,mm$ pitch).

Substituting Eq.~\ref{eq:N} into Eq.~\ref{eq:L}, we obtain the dependence
of the average network size on the PE clock frequency:
\begin{equation}
L_g=\sqrt{f_0WQ\left(\frac{\sigma D}{B_w\alpha}\right)}.\label{eq:L1}
\end{equation}
Notice that the parameters in the parentheses are beyond our control.
($D$ is a slow function of $N$ and can be considered a constant.) 

Combining Eq.~\ref{eq:X} and Eq.~\ref{eq:L1}, we discover that the average
``packing'' size of the interprocessor network again scales as the square root
of the performance of the hypercomputer:
\begin{equation}
L_g=\sqrt{\Theta W_0\left(\frac{\sigma D}{B_w\alpha}\right)}.\label{eq:L2}
\end{equation}
We would like to emphasize that Eq.~\ref{eq:L2} has been obtained
exclusively by considering the geometric volume necessary to contain
the passive interconnecting wires.

The ``packing'' size of the hypercomputer given by Equation~\ref{eq:L2} is
almost 9.8$\,m$.

\Section{Parallelism}
The ``well-balanced'' condition postulated in Sec.~\ref{sec:model}
imposes even stricter requirements on
the scaling of a hypercomputer.  The net effect of the geometry of
the system on the expected degree of parallelism will be discussed in
this section.

In a ``well balanced'' system, the number of thread contexts per PE
(or the amount of parallelism, $T$) must be large enough to
tolerate the round trip latency of a memory access measured in PE
clock cycles. The latency includes the signal propagation time
$\tau_p$, message processing overhead $\tau_n$, and memory response
time $\tau_m$:
\begin{equation}
T=\left(\tau_p+\tau_n+\tau_m\right)f_0 = 
\left(\frac{2LD}{c_s}+\frac{DC}{f_0}+\tau_m\right)f_0.\label{eq:TS}
\end{equation}
Here, $c_s$ is the signal propagation speed
(in copper, $c_s\approx9\cdot10^7\,m/s$), and $C$ is the number of PE cycles
required for message processing at one internal network node (we take
$C\sim10$, but believe that it may be as low as 1). It can be shown
that for the hypercomputer proposed above, the first term
dominates the other two. Indeed, $\tau_p\approx2.25\,\mu s$,
$\tau_n\approx5\,ns$ (the first and the second
terms in Eq.~\ref{eq:TS}), and $\tau_m\approx1\,ns$~\cite{itrs99}. For the
rest of our reasoning, we may safely assume that
\begin{equation}
T\approx Lf_0\left(2D/c_s\right). \label{eq:T}
\end{equation}

\begin{figure}[!tb]\centering
\epsfig{file=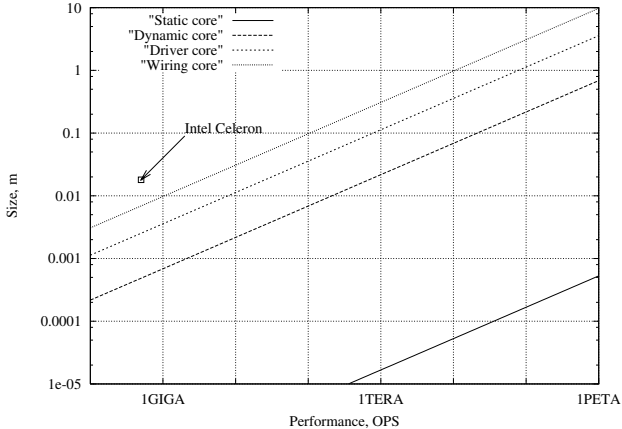,height=\columnwidth,angle=-90}
\caption{\label{pict:diam}The minimal diameter of the TVHC installation as a function of its performance.}
\end{figure}

The comparison of Eq.~\ref{eq:Lpower} and Eq.~\ref{eq:L2} with respect to
``our'' hypercomputer suggests (Figure~\ref{pict:diam}) that the
geometric considerations dominate the power-management
considerations, regardless of the performance of the
installation. Therefore, the study of the power consumption may be
safely omitted, and we can concentrate on the geometric term.

The combination of Eq.~\ref{eq:L2} and Eq.~\ref{eq:T} gives the dependence
of $T$ on the hypercomputer clock frequency and overall performance:

\begin{equation}
T=f_0\sqrt{\Theta}\left(\frac{2}{c_s}\sqrt{\frac{\sigma
D^3W_0}{B_w\alpha}}\right).\label{eq:T1}
\end{equation}
For the  TVHC, $T\sim70,000$.  As usual, the
factors collected in the parentheses are beyond our control.

\Section{Solutions}
An unpleasant consequence of the equation~\ref{eq:T1} is that the
amount of intrinsic parallelism required from an application in order
to be efficiently executed by a hypercomputer is proportional to the
clock frequency of the PE and to the square root of the overall
performance of the machine. This means that ``super-fast computing''
and ``high-performance computing'' are essentially synonyms for
``massively-parallel computing'', and as such cannot be considered
suitable for general-purpose applications with a purely random memory
access pattern.

A number of solutions may be suggested to this problem. One way to
circumvent the ``packing'' constraint is to use open-space optical
interconnects. For these kind of links, one can expect to have the
bandwidth $B_0\approx40\,Gbps$ per link, with signal propagation speed
$c_s=3\cdot10^8\,m/s$. An important property of an open-space network
is that the links can actually overlap. Therefore, the size of the
core will not be limited by volume anymore. Instead, it will be
limited by the area of the inner surface of the shell:
$$
L_g=\sqrt{4N\sigma_{\mathrm LE}/\pi}.
$$
Here, $\sigma_{\mathrm LE}$ is the footprint of a light
emitting element, for instance, vertical cavity surface emitting laser
(VCSEL). Assuming that the size of a VCSEL is $200\,\mu
m\times200\,\mu m$~\cite{lasermate01}, the diameter of the shell $L_g$ will be
$\approx2\,m$ --- a big improvement, compared to the ``copper''
shell. It is also worth mentioning that the static power dissipation
in an open-space network is zero, due to the absence of
wires. 

We could not find reliable information on the power consumption of
very high-speed VCSELs and photodiodes. An intelligent guess is that
at $40\,Gbps$, power required by a single emitter is
$\sim0.1\,mW$. Equation~\ref{eq:Ldriver} gives the size of the
``driver core'': $L_{dr}\approx3.3\,m$. As one can see, the ``driver''
shell becomes bigger than the ``packing'' shell and determines the
size of the TVHC. Once again, we would like to emphasize that we have
no solid numbers for very high-speed VCSELs, and the result of this
calculation must be considered exclusively as a rough estimate.

There exists at least yet another alternative to copper wires. They
can be replaced with high-speed ballistic high-$T_c$ superconductor
(HTSC) ceramic wires. HTSC wires promise high data transfer rates
($B_w\approx10\,Gbps$) and high signal propagation speed
($s_c\approx2\cdot10^8\,m/s$). These two factors together can reduce
the ``packing'' size and the degree of parallelism by 40\% and 60\%,
respectively. However, the ultimate cross-section of ceramic wires is
not know now, and this third factor may potentially undo the
improvement. There will be still at least some gain, unless the HTSC
wires are $6\cdot10^{-7}\,m^2$ in cross-section or thicker.

The biggest improvement that can be brought in by the HTSC wires is
the shrinkage of the ``driver'' core. Superconductor drivers may
consume as little as $10\,\mu W$ of power, compared to $20\,mW$ for
semiconductor drivers. This would reduce the size of the respective
core to $L_{dr}\approx0.5\,m$, which would allow us to totally exclude
it from the consideration.

Unfortunately, HTSC wires can operate only at the temperature of
liquid nitrogen and require deep refrigeration. The dissipated power
will be removed elsewhere (namely, at the nitrogen liquifier setup,
which may be located outside of the shell) and will not contribute to
the power balance of the core. However, the cryogenic infrastructure
may (and apparently will) inflate the effective cross-section $\sigma$
of the interconnects. The net effect of this inflation is not known
yet.

\setlength\textheight{4.25in}
\Section{Conclusion}
We have considered the parametric dependences of the geometric size of
a hypothetical petaflops-scale hypercomputer on the geometric size and
power properties of its interconnection network. We discovered that
the size of a hypercomputer with spherical arrangement of active
components (processing and switching elements and memories) scales as
the square root of the aggregate peak performance:
$L\sim\sqrt{\Theta}$. In order to sustain the execution rate, the
hypercomputer must be designed as a highly multithreaded machine.  As
such, it will be most suited for highly parallel applications. Even
though it may be possible to reduce the degree of parallelism by
optimizing the implementation of the network, it is questionable
whether a general-purpose application with purely random memory access
pattern can benefit from being executed by the hypercomputer.

\Section{Acknowledgments}

The author would like to thank Paul Ezust and Dan Stef\u{a}nescu (Suffolk
University) for useful discussions and help with the preparation of
the manuscript, and T.~Sterling (JPL), K.~Likharev (SUNY), and
P.~Bunyk (TRW) for the inspiration.
\bibliographystyle{latex8}
\bibliography{myself,petaflop,rsfq_rest,misc}

\end{document}